\documentclass[preprintnumbers, prd, onecolumn, showpacs,floatfix,preprintnumbers,
superscriptaddress, nofootinbib]{revtex4}
\usepackage{graphicx}
\usepackage{epsfig}
\usepackage{bm}
\usepackage{amssymb}
\usepackage{float}
\usepackage{amsmath}
\usepackage{subfigure}
\usepackage{dcolumn}
\usepackage{cancel}
\usepackage[colorlinks]{hyperref}
\usepackage[usenames,dvipsnames]{color}
\hypersetup{
     breaklinks=true,
    pdfstartview={FitH},    
    colorlinks=true,       
    linkcolor=blue,          
    citecolor=red,        
    filecolor=magenta,      
    urlcolor=blue,           
    anchorcolor=green,      
    linktocpage=true
}

\def\doi{http://doi.org}

\begin{document}
\title{Constraints on a generalized deceleration parameter from cosmic chronometers}
\author{Abdulla Al Mamon}
\email{abdulla.physics@gmail.com}
\affiliation{Department of Mathematics, Jadavpur University, Kolkata-700032, India}
\affiliation{Manipal Centre for Natural Sciences, Manipal University, Manipal-576104, India}
\newcommand{\be}{\begin{equation}}
\newcommand{\ee}{\end{equation}}
\newcommand{\bea}{\begin{eqnarray}}
\newcommand{\eea}{\end{eqnarray}}
\newcommand{\bc}{\begin{center}}
\newcommand{\ec}{\end{center}}
\begin{abstract}
In this paper, we have proposed a generalized parametrization for the deceleration parameter $q$ in order to study the evolutionary history of the universe. We have shown that the proposed model can reproduce three well known $q$-parametrized models for some specific values of the model parameter $\alpha$. We have used the latest compilation of the Hubble parameter measurements obtained from the cosmic chronometer (CC) method (in combination with the local value of the Hubble constant $H_{0}$) and the Type Ia supernova (SNIa) data to place constraints on the parameters of the model for different values of $\alpha$. We have found that the resulting constraints on the deceleration parameter and the dark energy equation of state support the $\Lambda$CDM model within $1\sigma$ confidence level at the present epoch. 
\end{abstract} 

\maketitle
\section{Introduction}
The late-time accelerated expansion phase of the universe is one of the biggest challenges in modern cosmology. A large number of cosmological observations \citep{riess1998,perl1999,teg2004,eisen2005,seljak2005,koma2011,hin2013,pl2014a,pl2015,
pl2016a,pl2016b,pl2016c} have strongly confirmed the current accelerated expansion phase of the universe. All of these observations also strongly suggest that the observed cosmic acceleration is rather a recent phenomenon and the universe was decelerating in the past. In the literature, the most accepted idea is that an exotic component of the matter sector with large negative pressure, dubbed as ``dark energy'', is responsible for this acceleration. According to recent observational data, almost $68.3\%$ of the total energy budget of the universe at the current epoch is filled with dark energy while $26.8\%$ with dark matter and remaining $4.9\%$ being the baryonic matter and radiation \citep{pl2014b,pl2014c,pl2014d}. However, understanding the origin and nature of the dark sectors (dark energy and dark matter) is still a mystery. Different possibilities have been explored in order to find an explanation of the late-time observed cosmic acceleration and for some excellent reviews on dark energy models, one can refer to \cite{vs2000,peeb2003,bamba2012,cope2006,martin2008}. The simplest and consistent with most of the observations is the $\Lambda$CDM (cosmological constant $\Lambda$ with pressureless cold dark mater) model where the constant vacuum energy density serves as the dark energy candidate. However, the models based upon cosmological constant suffer from two serious drawbacks, namely, the {\it fine tuning} and the cosmological {\it coincidence} problems \citep{wein1989,Steinhardt1999}. Therefore, most of the recent research is aimed towards finding a suitable cosmologically viable model of dark energy.\\
\par Parameterization of the deceleration parameter $q$ is a useful tool towards a more complete characterization of the evolution history of the universe. Several well-known parameterizations for the deceleration parameter have
been proposed so far (for a review, see \cite{turner2002,riess2004,gong2006,gong2007,chuna2008,chuna2009,xu2008,xu2009,santos2011,
nair2012,del2012,aksaru2014,mamon2016a,mamon2016b}).\\
\par In the next section, we have proposed a generalized model for the deceleration parameter in order to explore late-stage evolution of the universe. One of the main properties of this model is that it can reproduce three popular $q$-parametrized models for some specific values of the model parameter. The constraints on the model parameters of our model have also been obtained using the SNIa, CC and $H_{0}$ datasets. We have shown that the present model describes the evolution of the universe from an early decelerated phase ($q>0$) to an accelerated phase ($q<0$) at the present epoch for the combined dataset (SNIa+CC+$H_{0}$).\\
\par The organization of the paper is as follows. In section \ref{sec2}, we have described the phenomenological model considered here. In section \ref{data}, we have described the observational datasets (including their analysis methods) adopted in this work and discussed the results in section \ref{result}. Finally, in section \ref{conclusion}, we have summarized the conclusions of this work.
\section{Theoretical model}\label{sec2}
The action for a scalar field $\phi$ and the Einstein-Hilbert term is 
described as 
\be\label{action}
S = \int d^4 x \sqrt{-g} \left( \frac{R}{16\pi G} 
 + \frac{1}{2}\partial_\mu \phi \partial^\mu \phi - V(\phi) \right) + S_m 
\ee
where $g$ is the determinant of the metric $g_{\mu\nu}$, $R$ is the scalar curvature, $V(\phi)$ is the scalar potential and $S_m$ is the action of the background matter which is considered as pressureless ($p_{m}=0$) perfect fluid. We have assumed a homogeneous, isotropic and spatially flat Friedmann-Robertson-Walker (FRW) universe which is characterized by the following line element
\begin{equation}
ds^{2} = dt^{2} - a^{2}(t)[dr^{2} + r^{2}(d{\theta}^{2} + \text{sin}^{2}\theta d{\phi}^{2})]
\label{eq:2.2}
\end{equation}
where, $a(t)$ is the scale factor of the universe which is scaled to be unity at the present epoch. The Einstein field equations corresponding to action (\ref{action}) for a spatially flat FRW geometry can be obtained as
\be\label{fe1}
3H^{2}=8\pi G(\rho_{m}+\rho_{de})
\ee
\be\label{fe2}
2{\dot{H}} + 3H^{2}=-8\pi G p_{de}
\ee
where $H=\frac{\dot{a}}{a}$ is the Hubble parameter and the dot implies derivative with respect to the cosmic time $t$. Here, $\rho_{m}$ represents the energy density of the dust matter while $\rho_{de}=\frac{1}{2}{\dot{\phi}}^{2} + V(\phi)$ and $p_{de}=\frac{1}{2}{\dot{\phi}}^{2} - V(\phi)$ represent the energy density and pressure of the dark energy respectively.\\
Also, the conservation equations for the dark energy and matter field are
\be
{\dot{\rho}}_{de}+3H(\rho_{de}+p_{de})=0 
\ee
\be
{\dot{\rho}}_{m}+3H\rho_{m}=0  
\ee
\par In cosmology, the deceleration parameter $q$ plays an important role in describing the nature of the expansion of the universe. Usually, it is parametrized as
\be\label{eqparaqm}
q(z)=q_{0} + q_{1}F(z)
\ee
where $q_{0}$, $q_{1}$ are constants and $F(z)$ is a function of the redshift $z$. In fact, various functional forms of $F(z)$ have been proposed in the literature \citep{turner2002,riess2004,gong2006,gong2007,chuna2008,chuna2009,xu2008,xu2009,santos2011,
nair2012,del2012,aksaru2014,mamon2016a,mamon2016b} which can provide a satisfactory solution to some of the cosmological problems. Few popular $q(z)$ models are: (i)~$q(z)=q_{0}+q_{1}z$, (ii)~$q(z)=q_{0}+q_{1}{\rm ln}(1+z)$, (iii)~$q(z)=q_{0}+q_{1}{\Big(\frac{z}{1+z}\Big)}$ and many more. However, some of these parametrizations diverge at $z\rightarrow -1$ and others are valid for $z<<1$ only. Recently, Mamon and Das \citep{mamon2016a,mamon2016b} proposed divergence-free parametrizations of $q(z)$ to investigate the whole expansion history of the universe. They have shown that such models are more consistent with the recent observational data for some restrictions on model parameters. Therefore, search is still on for an appropriate functional form of $q(z)$ that will describe the evolutionary history of the universe and fit well in dealing with cosmological challenges.\\
\par In order to close the system of equations as well as to extend the $q$-parametrizations above, in the present work, we propose
\be\label{ans}
q(z)=q_{0}-q_{1}{\left[\frac{(1+z)^{-\alpha}-1}{\alpha}\right]}
\ee
where $q_{0}$, $q_{1}$ and $\alpha$ are arbitrary model parameters. It deserves to mention here that $q_{0}$ denotes the present value of $q$, and $q_{1}$ denotes the derivative of $q$ with respect to the redshift $z$.  From equation (\ref{ans}), it is straightforward to show that the deceleration parameters given by examples (i)-(iii) are fully recovered in the following limits:
\bea\label{eqlimit}
{q(z)} = \left\{\begin{array}{ll} q_{0}+q_{1}z,&$for$\ \alpha= -1 \\\\
q_0 +q_{1}\ln(1+z),\ \ \ \ \ \ \ \ \ \ &$for$\
\alpha\rightarrow 0 \\\\
q_{0}+q_{1}{\left(\frac{z}{1+z}\right)},&$for$\ \alpha=+1
\end{array}\right..
\eea
Thus, the new parametrization of $q(z)$ covers a wide range of popular theoretical models (for $\alpha=\pm 1$ and $\alpha\rightarrow 0$) and also provides a new range of cosmological solutions in a more general framework for $\alpha \neq -1$, $0$, $+1$. It is important to note that for the limit $\alpha \rightarrow 0$, we have used the equality, ${\rm lim}_{\epsilon\rightarrow 0}{\Big[\frac{x^{\epsilon}-1}{\epsilon}\Big]}={\rm ln}x$. It deserves mention here that our parametrization is not valid at $\alpha=0$. It is worth noting here that the parametric form of $q(z)$, given by equation (\ref{ans}), is inspired from the generalized parametrization of the equation of state parameter for dark energy \cite{motibeosp}.\\
\par The deceleration parameter and the Hubble parameter are related by the following equation
\be\label{eqhqlog}
H(z)=H_{0}{\rm exp}{\left(\int^{z}_{0} \frac{1+q(x)}{1+x}dx\right)}
\ee
where $H_{0}$ indicates the present value of the Hubble parameter. Using equations (\ref{ans}) and (\ref{eqhqlog}), the Friedmann equation for our generalized model can be obtained as
\be\label{eqhz}
H(z)=H_{0}(1+z)^{(1+q_{0}+\frac{q_{1}}{\alpha})} {\rm exp}{\Big[\frac{q_{1}}{\alpha^{2}}\left\{(1+z)^{-\alpha}-1\right\}\Big]}
\ee
It should be noted that the present mechanism can also be treated as parameterization of the Hubble parameter which can be constrained from observational Hubble data directly.\\
For this model, we obtained the expression for the equation of state parameter as
\be
w_{de}=\frac{p_{de}}{\rho_{de}}=\frac{2q(z)-1}{3-3\Omega_{m0}(1+z)^{1-2q_{0}-\frac{2q_{1}}{\alpha}}{\rm exp}{\Big[\frac{2q_{1}}{\alpha^{2}}\left\{1-(1+z)^{-\alpha}\right\}\Big]}} 
\ee
where $\Omega_{m0}$ indicates the present value of the density parameter of the matter component.\\
\par The behavior and the main cosmological characteristics of the model given in equation (\ref{ans}) strongly depend on the model parameters ($q_{0}$, $q_{1}$ and $\alpha$). In the next section, using different observational datasets, we have constrained the model parameters ($q_{0}$, $q_{1}$) for some specific values of $\alpha$. In this work, we have worked with values ($\alpha \neq -1, 0, +1$) as these particular $q(z)$ models given in equation (\ref{eqlimit}) has been studied extensively.
\section{Data analysis methods}\label{data}
In this section, we have briefly discussed the datasets considered in our analysis, namely Type Ia Supernova (SNIa) and new dataset of Hubble parameter obtained with the cosmic chronometer (CC) method, in combination with the local value of $H_{0}$. In the following, we have described how these datasets are included into the $\chi^{2}$ analysis.
\subsection{Cosmic chronometer dataset and local value of the Hubble parameter}
The {\it cosmic chronometers} (CC) approach was first introduced by \cite{jim2002} to measure $H(z)$. It uses the relative ages of the most massive and passively evolving galaxies to measure $\frac{dz}{dt}$, from which $H(z)$ is deduced. In this work, we have used the latest updated list of Hubble parameter dataset \citep{simon2005,stern2010,zhang2014,more2015,more2016} obtained through the CC approach, comprising $30$ measurements spanning the redshift range $0 < z < 2$, recently compiled in \citep{more2015,more2016}. In our analysis, we have also included the recently measured local value of Hubble parameter given by Planck analysis ($H_{0}= 67.3 \pm 1.2$ km/s/Mpc) \citep{pl2014b}. We have obtained the constraints on the model parameters by minimizing the following $\chi^2$ function 
\be
\chi^2_{H} = \sum^{30}_{i=1}\frac{[{H}^{obs}(z_{i}) - {H}^{th}(z_{i},H_{0},\theta)]^2}{\sigma^2_{H}(z_{i})}, 
\ee
where ${H}^{obs}$ and ${H}^{th}$ are the observed and theoretical values of the Hubble parameter respectively. Also, $\sigma_{H}$ represents the uncertainty associated
with each measurement of $H$ and $\theta$ denotes the model parameter.
\subsection{Type Ia Supernova dataset}
Data from the Type Ia Supernova was the first indication for cosmic acceleration and it is very useful to test the cosmological models. So, along with the CC dataset, we have also utilized the distance modulus data sample of the recent ``joint light curve" (JLA) analysis in the present work \citep{Betoule2014}. For this dataset, the $\chi^{2}$ is defined as (see \cite{fmr2013} for more details)
\be
\chi^{2}_{SNIa}=A(\theta)-\frac{{B}^2(\theta)}{C}-\frac{2{\rm ln}10}{5C}B(\theta)-Q
\ee
with
\bea
A(\theta)=\sum_{\alpha,\beta}(\mu^{th}-\mu^{obs})_{\alpha}({\cal C}ov)^{-1}_{\alpha\beta}(\mu^{th}-\mu^{obs})_{\beta},\\
B(\theta)=\sum_{\alpha}(\mu^{th}-\mu^{obs})_{\alpha}\sum_{\beta}({\cal C}ov)^{-1}_{\alpha\beta},~~~~~~~~~~\\
C=\sum_{\alpha,\beta}({\cal C}ov)^{-1}_{\alpha\beta}~~~~~~~~~~~~~~~~~~~~~~~~~~~~~~~~~
\eea
Here, $``{\cal C}ov"$ is the covariance matrix of the data sample, $Q$ is a constant that does not depend on the model parameter $\theta$ and hence has been ignored. Here, $\mu^{obs}$ denotes the observed distance modulus, while $\mu^{th}$ is for the theoretical one, which is defined as
\be
\mu^{th}=5{\rm log}_{10}{\Big[\frac{d_{L}(z)}{1{\rm Mpc}}\Big]}+ 25
\ee
where, 
\be
d_{L}(z)=\frac{(1+z)}{H_{0}}\int^{z}_{0}\frac{dx}{h(x)}~~~{\rm and}~~~h=\frac{H}{H_{0}}.
\ee
\par Since the SNIa and CC are effectively independent observations, we can combine these datasets by adding together the $\chi^2$ functions. Thus, the combined $\chi^2$ can be expressed as
\be
\chi^{2}=\chi^{2}_{SNIa}+\chi^{2}_{H}. 
\ee
For the combined dataset (SNIa+CC+$H_{0}$), we estimate the best-fit values of the model parameters by minimizing $\chi^2$. Then, we use the maximum likelihood method and take the combined likelihood function as ${\cal L}={\rm e}^{-\frac{\chi^{2}}{2}}$. The best-fit parameter values $\theta^{*}$ are those that minimize the likelihood function
\be
{\cal L}(\theta^{*})={\rm exp}{\Big[-\frac{\chi^2(\theta^{*})}{2}\Big]}
\ee
We can now plot the contours for different confidence levels, e.g., $1\sigma~(68.3\%)$ and $2\sigma~(95.4\%)$. It is notable that the confidence levels $1\sigma$ and $2\sigma$ are taken proportional to $\bigtriangleup \chi^2 =2.3$ and $\bigtriangleup \chi^2 =6.17$ respectively, where $\bigtriangleup \chi^2 =\chi^2 (\theta) - \chi^{2}_{m}(\theta^{*})$ and  $\chi^{2}_{m}$ is the minimum value of $\chi^{2}$. An important quantity which could be used for data fitting process is \cite{refapss1,refepjc2,refijmpd3}
\be
{\bar{\chi}}^2=\frac{\chi^{2}_{m}}{N_{dof}} 
\ee
where, the subscript `dof' is abbreviation of {\it degree of freedom}, and $N_{dof}$ is defined as the difference between all observational data sources and the amount of free parameters. If ${\bar{\chi}}^{2} \le 1$, then the fit is good and the data are consistent with the considered model.
\section{Results of the data analysis}\label{result} 
\begin{figure*}
\resizebox{10cm}{!}{\rotatebox{0}{\includegraphics{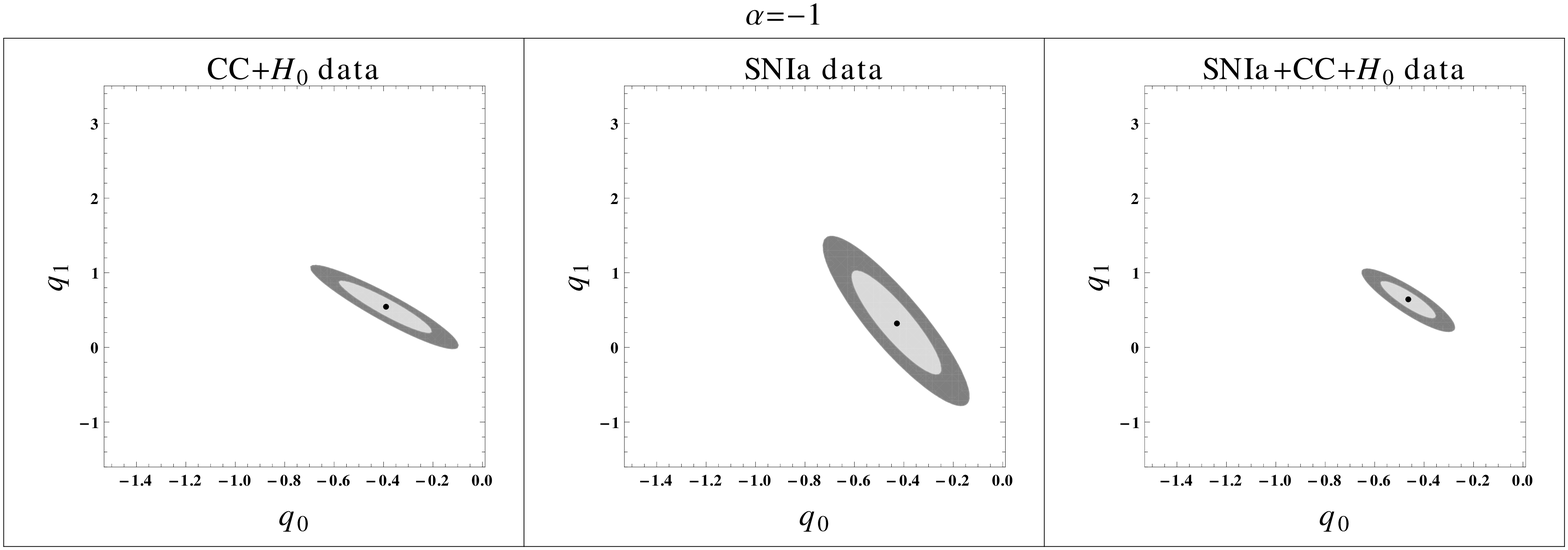}}}\\
\vspace{2mm}
\resizebox{10cm}{!}{\rotatebox{0}{\includegraphics{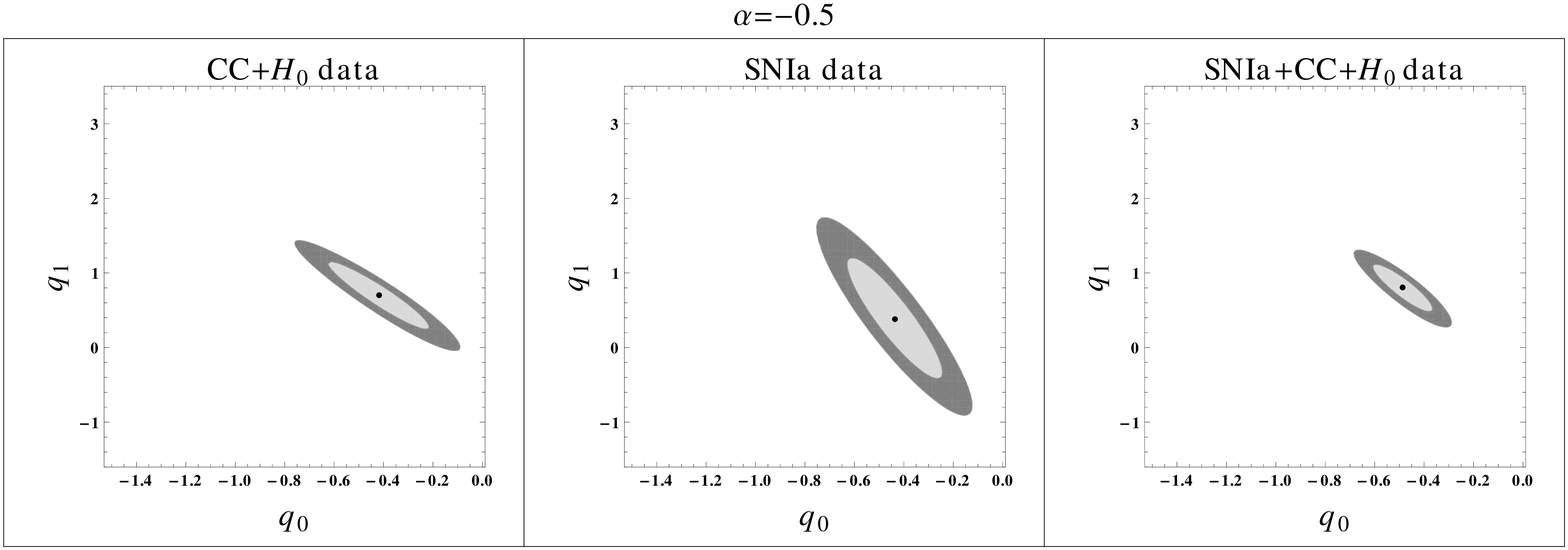}}}\\
\vspace{2mm}
\resizebox{10cm}{!}{\rotatebox{0}{\includegraphics{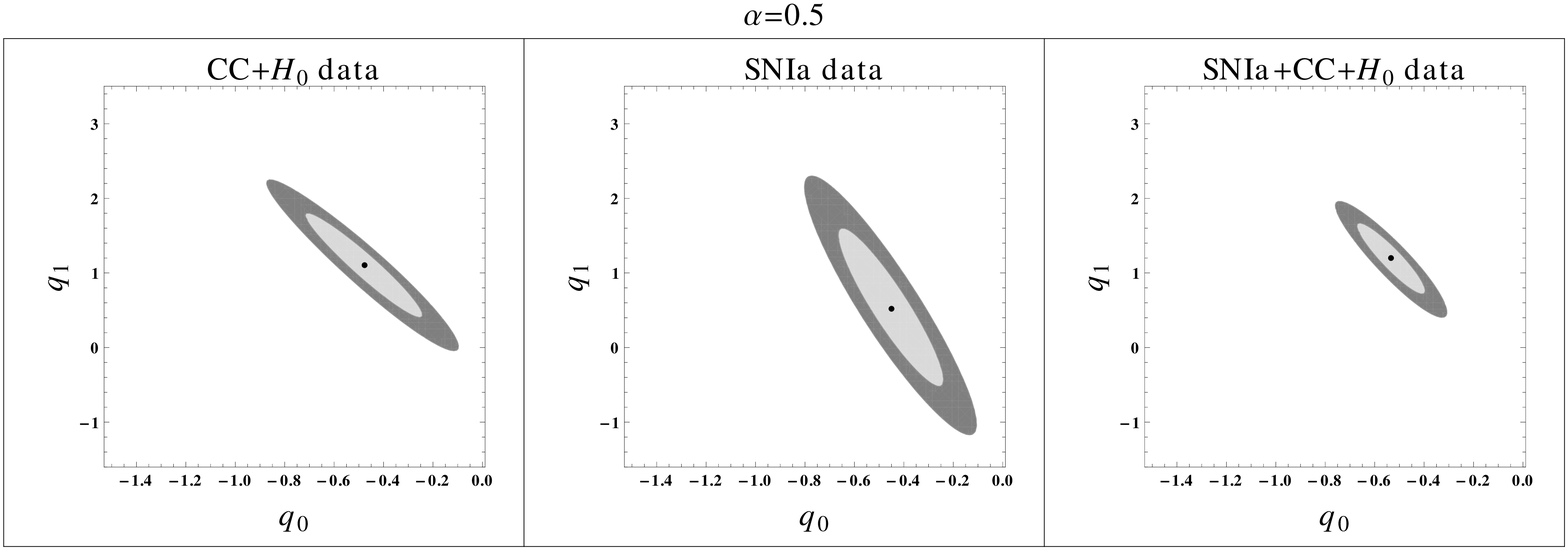}}}\\
\vspace{2mm}
\resizebox{10cm}{!}{\rotatebox{0}{\includegraphics{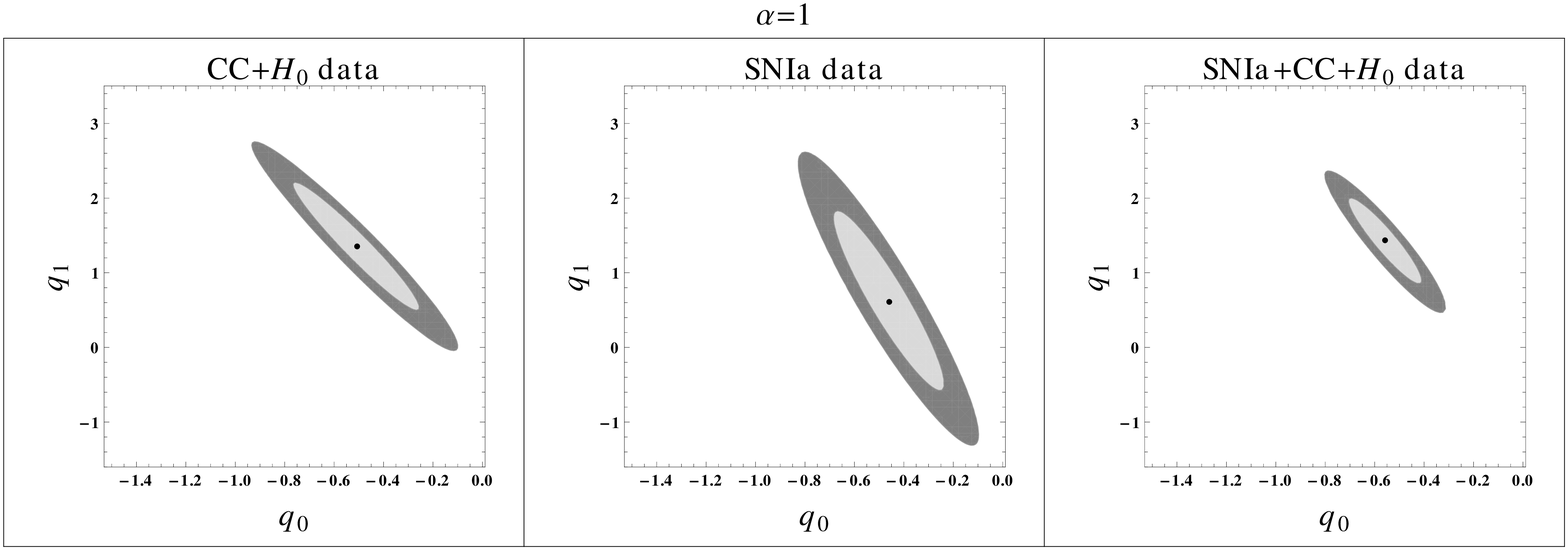}}}\\
\vspace{2mm}
\resizebox{10cm}{!}{\rotatebox{0}{\includegraphics{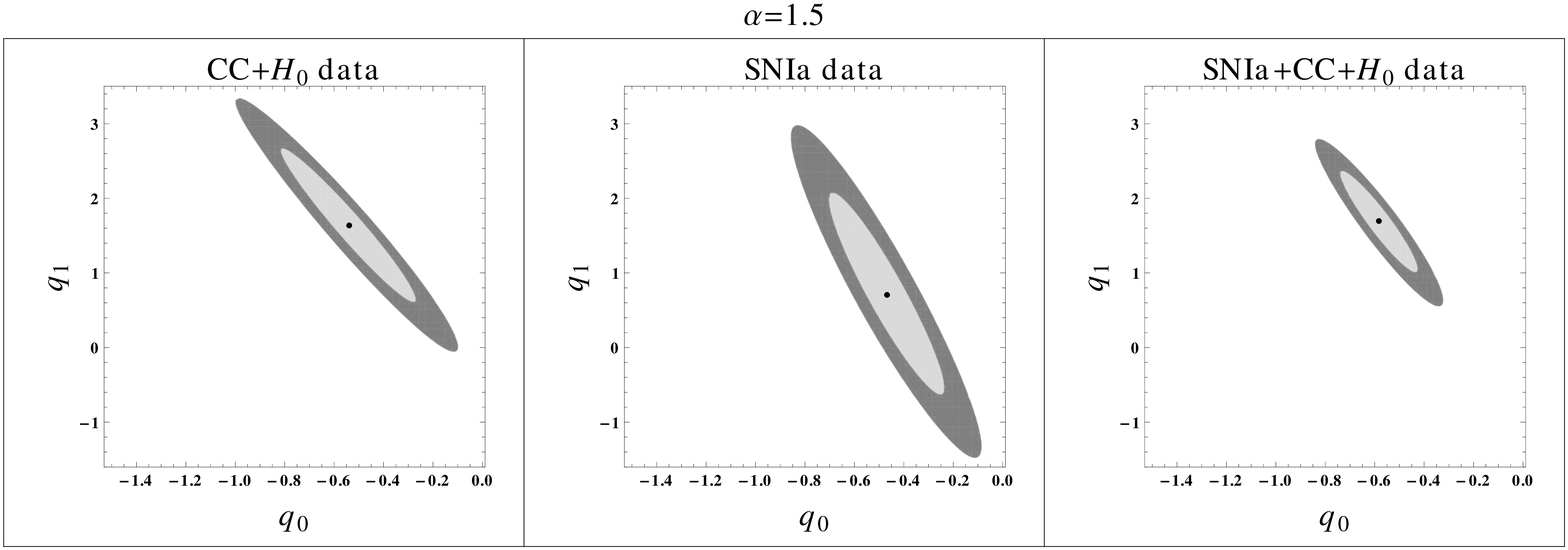}}}
\caption{This figure shows the $1\sigma$ and $2\sigma$ confidence contours for different choices of $\alpha$ and using the various datasets (CC+$H_{0}$, SNIa and SNIa+CC+$H_{0}$), as indicated in each panel. In each panel, the large dot represents the best-fit values of the pair ($q_{0},q_{1}$).}
  \label{fig1}
\end{figure*}

\begin{figure*}
\resizebox{12cm}{!}{\rotatebox{0}{\includegraphics{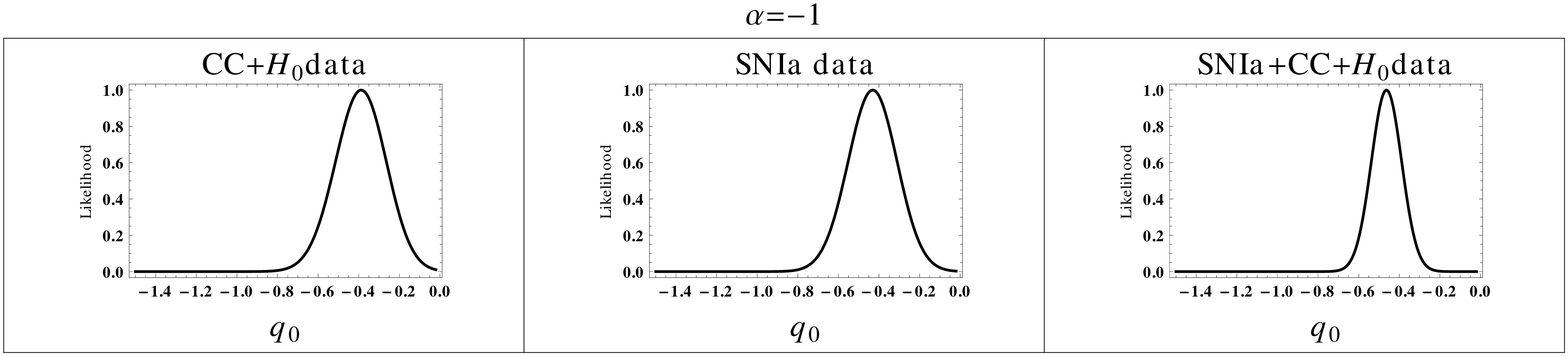}}}\\
\vspace{3mm}
\resizebox{12cm}{!}{\rotatebox{0}{\includegraphics{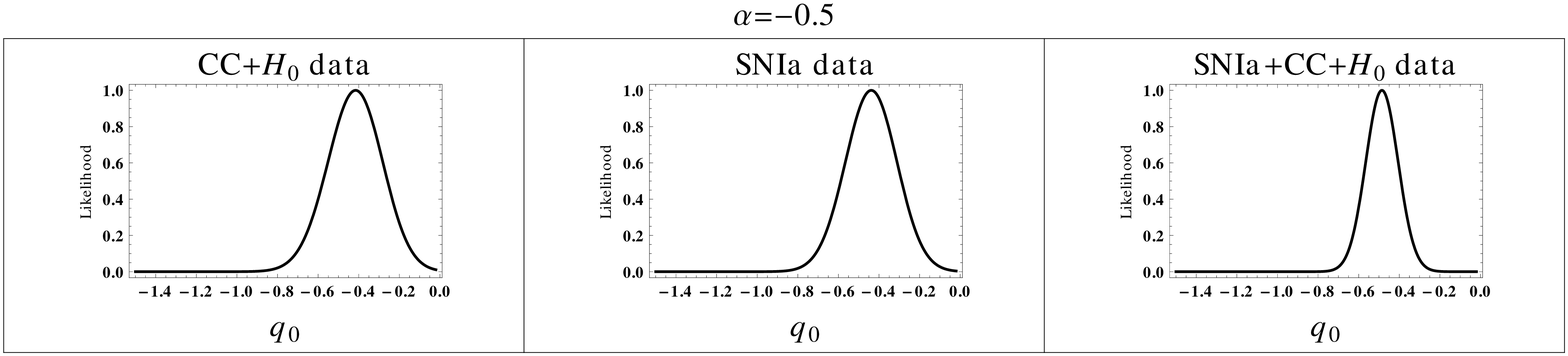}}}\\
\vspace{3mm}
\resizebox{12cm}{!}{\rotatebox{0}{\includegraphics{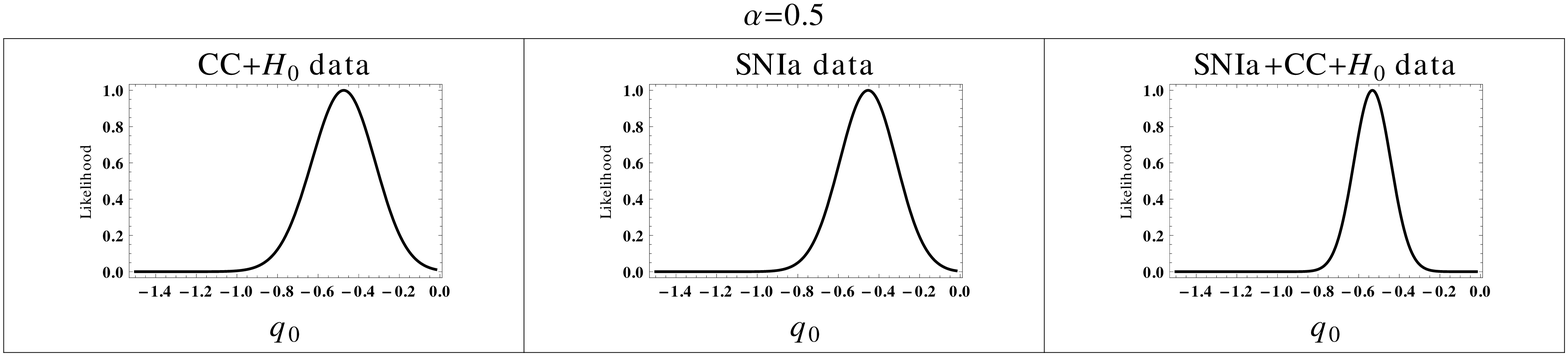}}}\\
\vspace{3mm}
\resizebox{12cm}{!}{\rotatebox{0}{\includegraphics{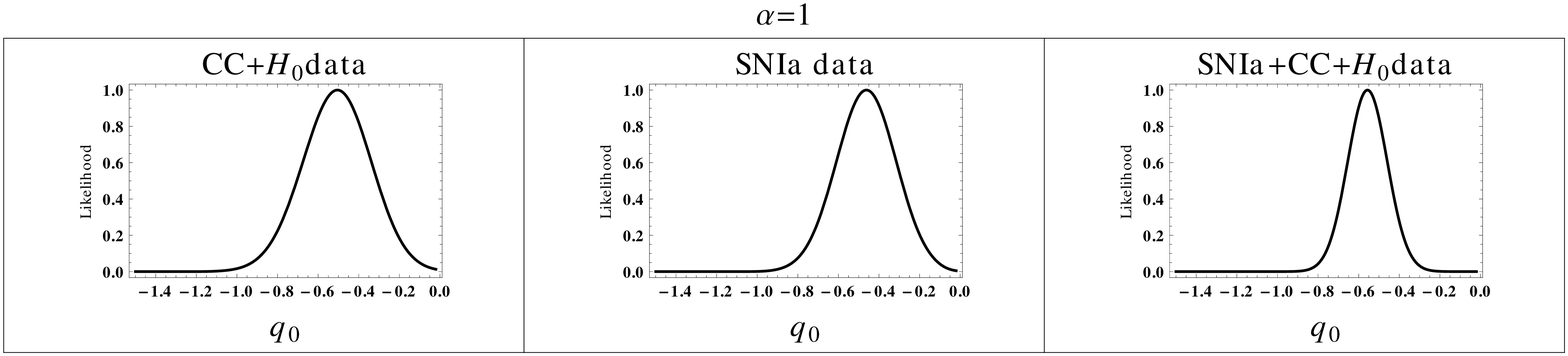}}}\\
\vspace{3mm}
\resizebox{12cm}{!}{\rotatebox{0}{\includegraphics{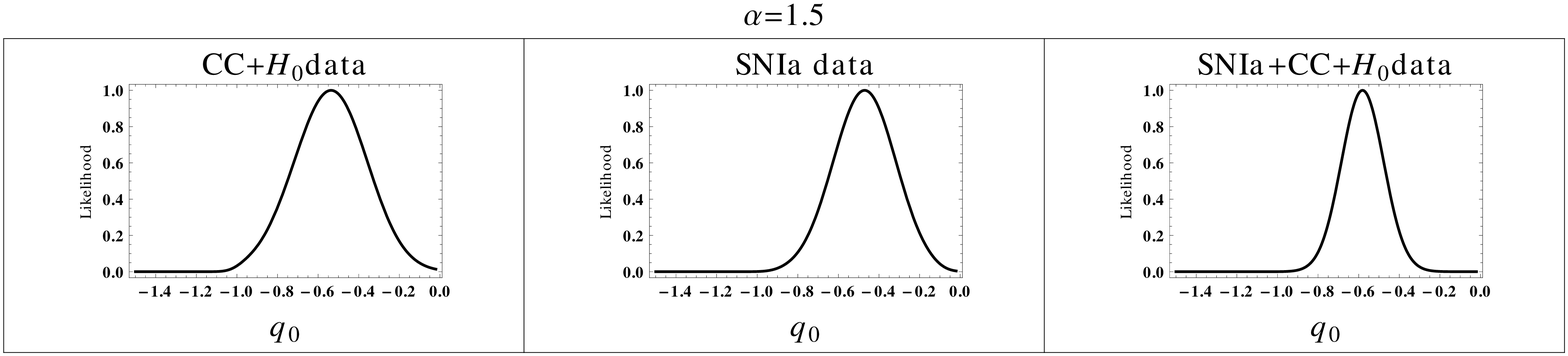}}}
\caption{This figure shows the marginalized likelihood function vs. $q_{0}$ for different choices of $\alpha$ using the various datasets, as indicated in each panel.}
  \label{fig2}
\end{figure*}

\begin{figure*}
\resizebox{12cm}{!}{\rotatebox{0}{\includegraphics{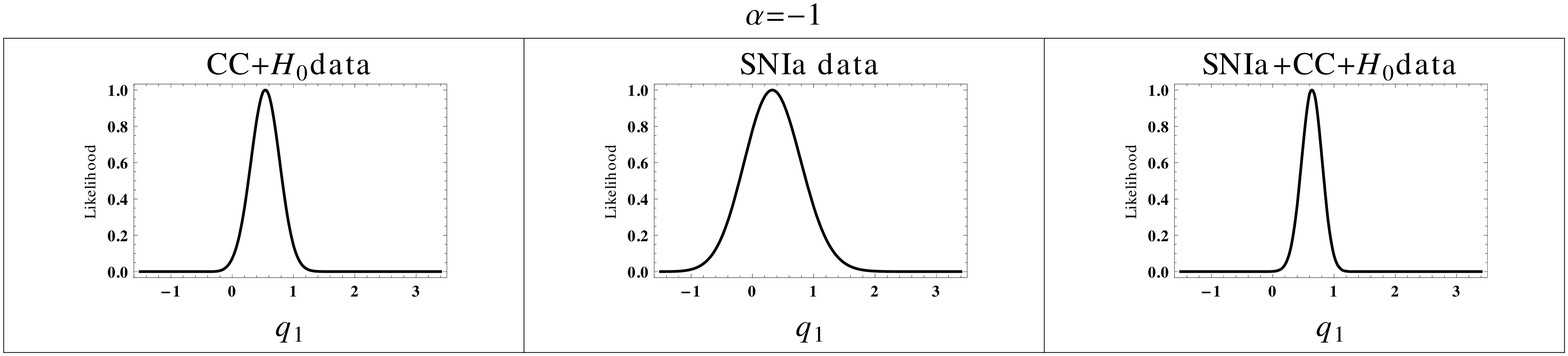}}}\\
\vspace{3mm}
\resizebox{12cm}{!}{\rotatebox{0}{\includegraphics{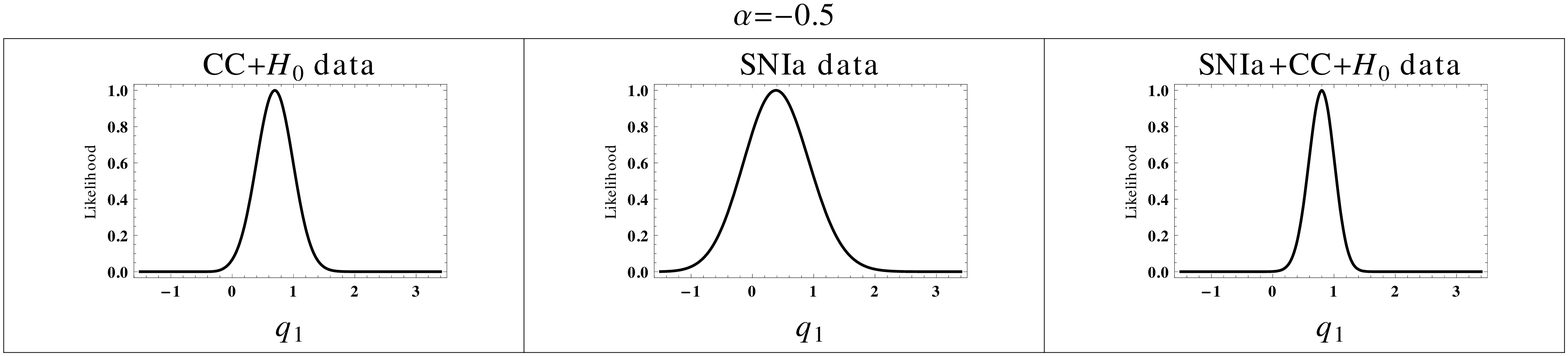}}}\\
\vspace{3mm}
\resizebox{12cm}{!}{\rotatebox{0}{\includegraphics{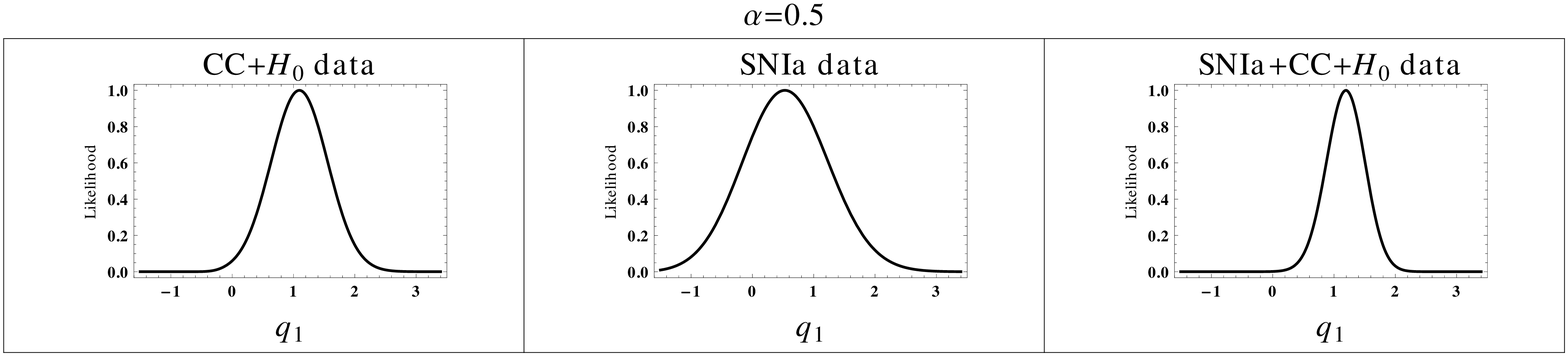}}}\\
\vspace{3mm}
\resizebox{12cm}{!}{\rotatebox{0}{\includegraphics{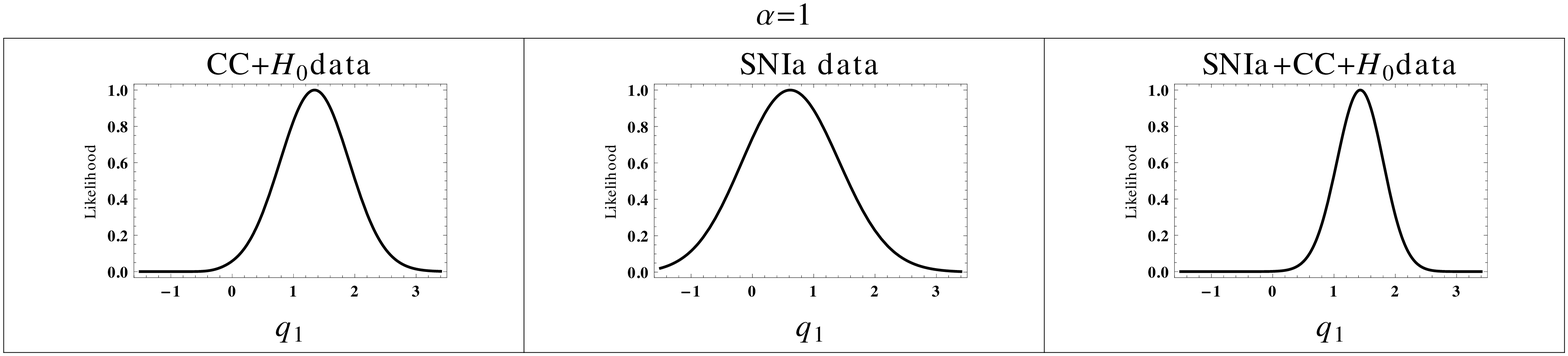}}}\\
\vspace{3mm}
\resizebox{12cm}{!}{\rotatebox{0}{\includegraphics{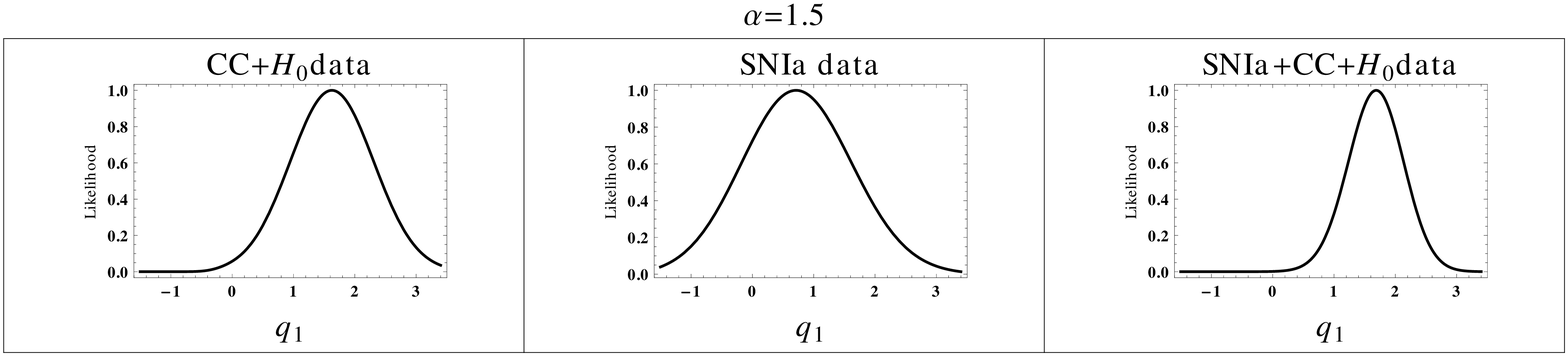}}}
\caption{This figure shows the marginalized likelihood function vs. $q_{1}$ for different choices of $\alpha$ using the various datasets, as indicated in each panel.}
  \label{fig3}
\end{figure*}

Figure \ref{fig1} shows the $1\sigma$ and $2\sigma$ contours in $q_{0}-q_{1}$ plane for the generalized $q$-parametrization given by equation (\ref{ans}). We have found that the best-fit values of the free parameters ($q_{0}$ and $q_{1}$) for the CC+$H_{0}$, SNIa and SNIa+CC+$H_{0}$ datasets are well fitted in the $1\sigma$ confidence contour. We have also found that the constraints obtained on the parameter values by the joint analysis (SNIa+CC+$H_{0}$) are very tight as compared to the constraints obtain from the CC+$H_{0}$
dataset and the SNIa dataset independently. On the other hand, figure \ref{fig2} and \ref{fig3} show the plots of the marginalized likelihood as functions of the model parameters $q_{0}$ and $q_{1}$ respectively. It is observed from the likelihood plots that the likelihood functions are well fitted to a Gaussian distribution function for each dataset. The corresponding constraints on model parameters are summarized in table \ref{table1}, \ref{table2} and \ref{table3}.\\
\par
\begin{figure*}
\resizebox{4.2cm}{!}{\rotatebox{0}{\includegraphics{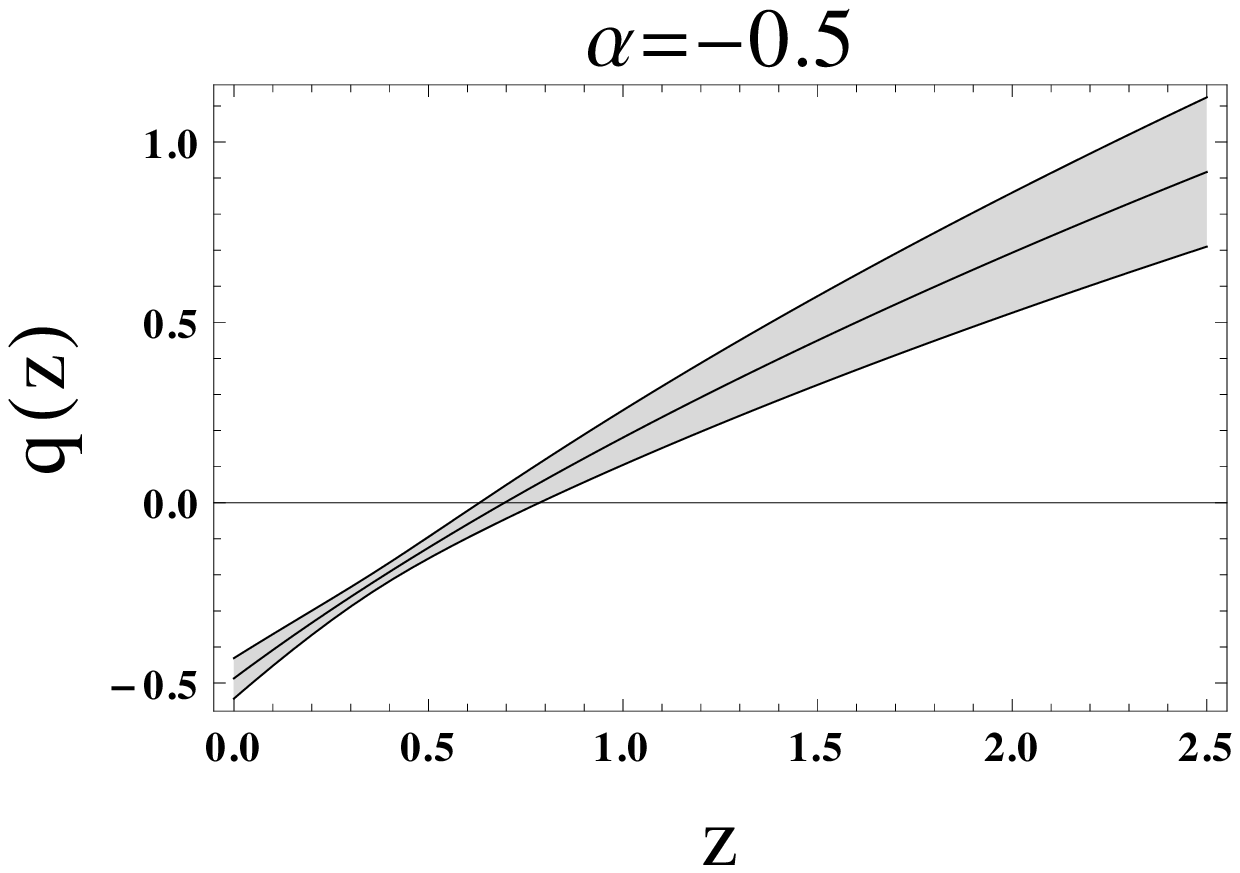}}}\hspace{3mm}
\resizebox{4.2cm}{!}{\rotatebox{0}{\includegraphics{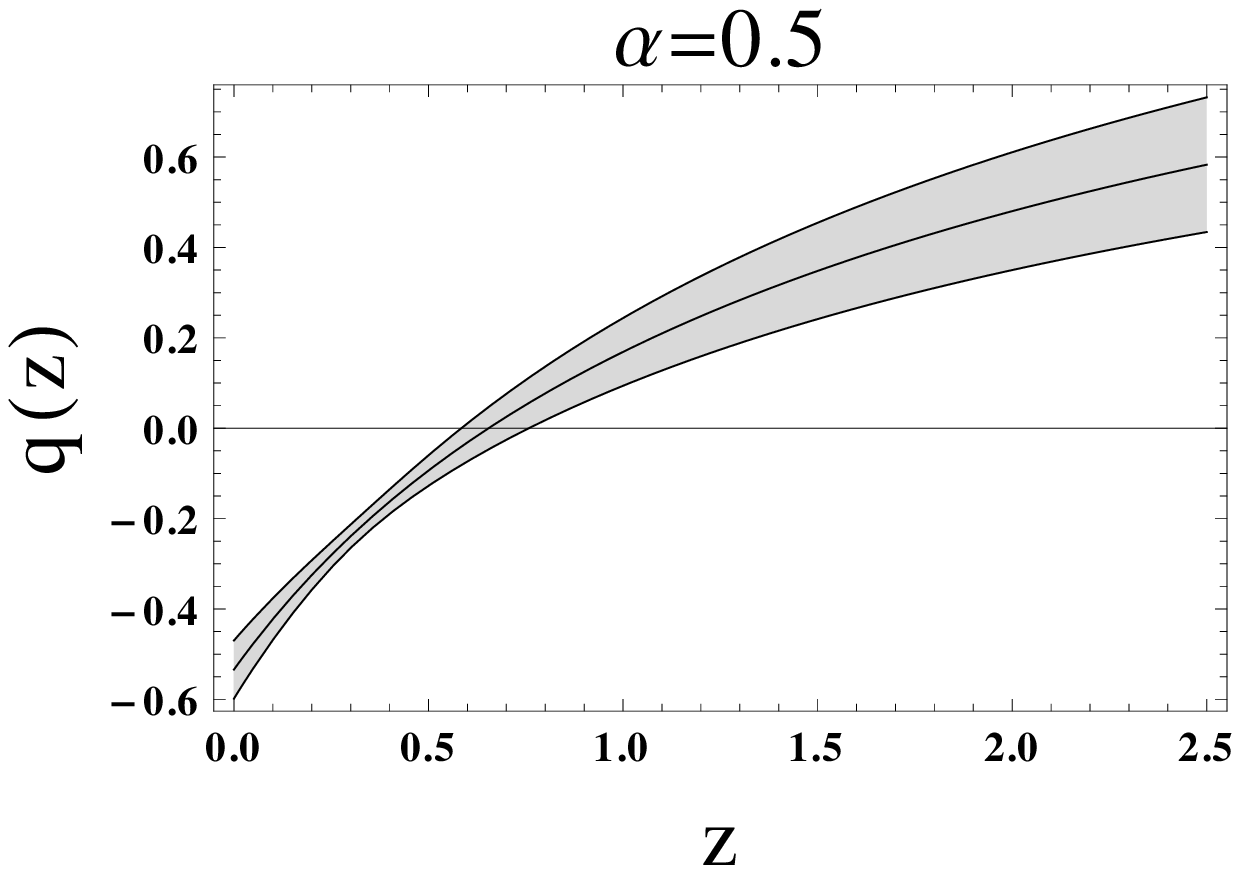}}}\hspace{3mm}
\resizebox{4.2cm}{!}{\rotatebox{0}{\includegraphics{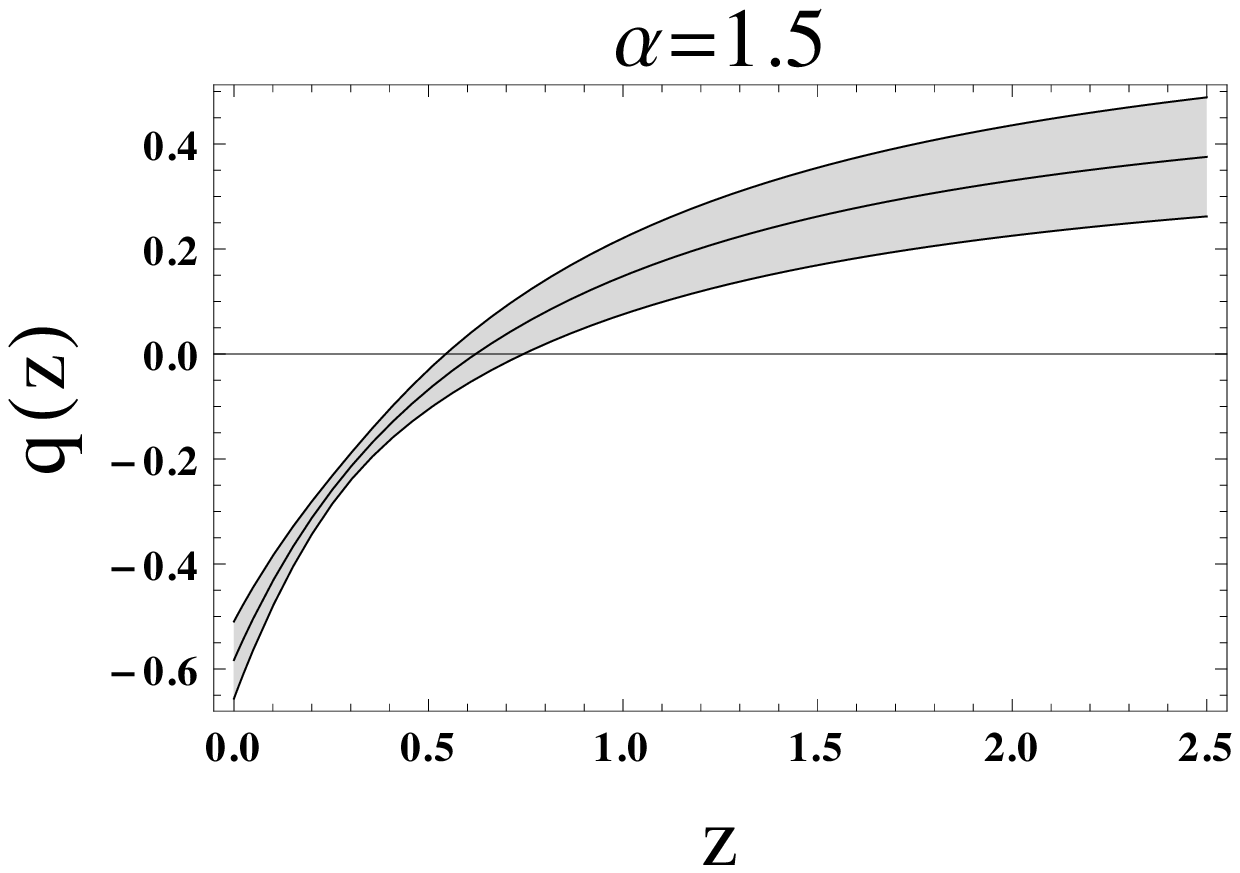}}}%
\caption{The evolution of deceleration parameter $q(z)$ is shown for the SNIa+CC+$H_{0}$ dataset by considering different values of $\alpha$, as indicated in each panel. In each panel, the central thin line and the light gray region represent the best fit curve and the $1\sigma$ confidence level respectively.}
  \label{figq}
\end{figure*}

\begin{figure*}
\resizebox{4.2cm}{!}{\rotatebox{0}{\includegraphics{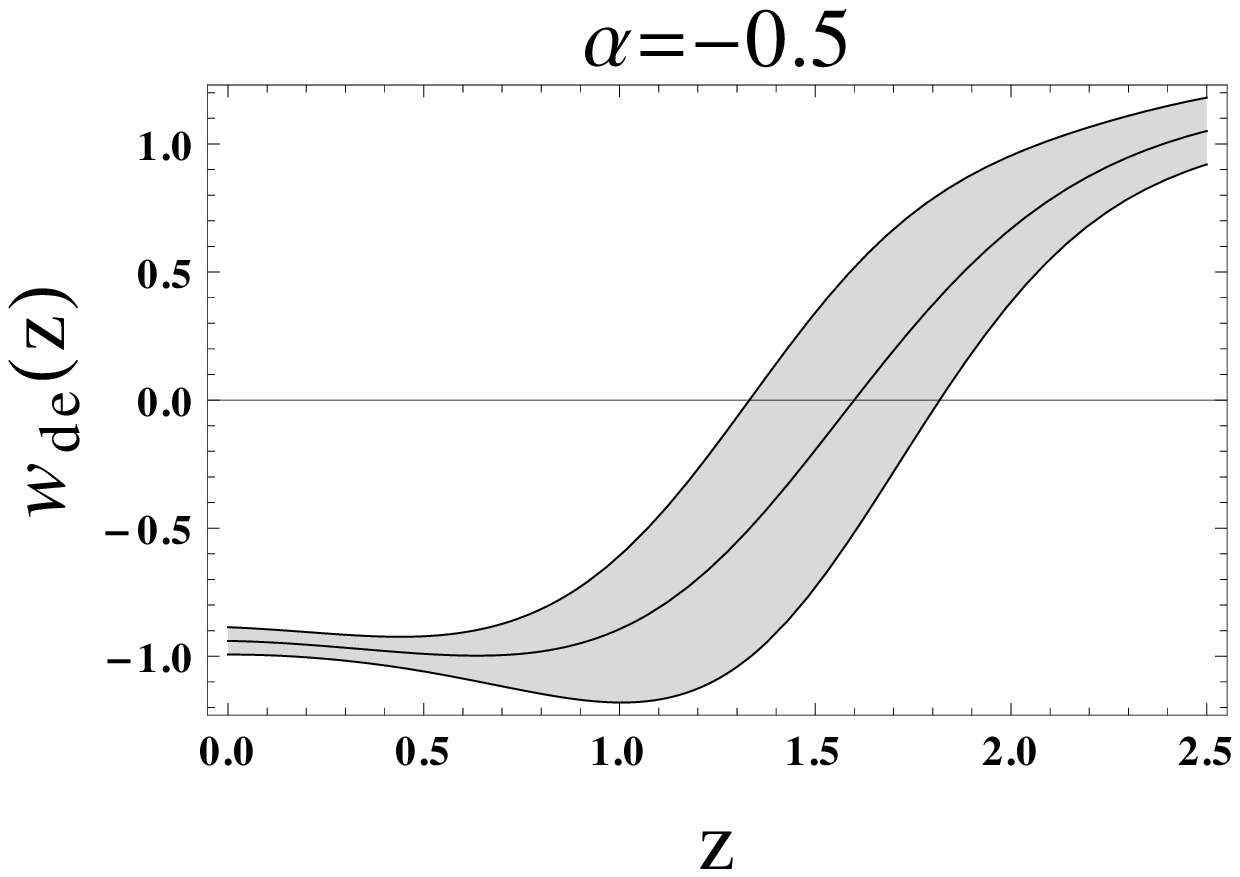}}}\hspace{3mm}
\resizebox{4.2cm}{!}{\rotatebox{0}{\includegraphics{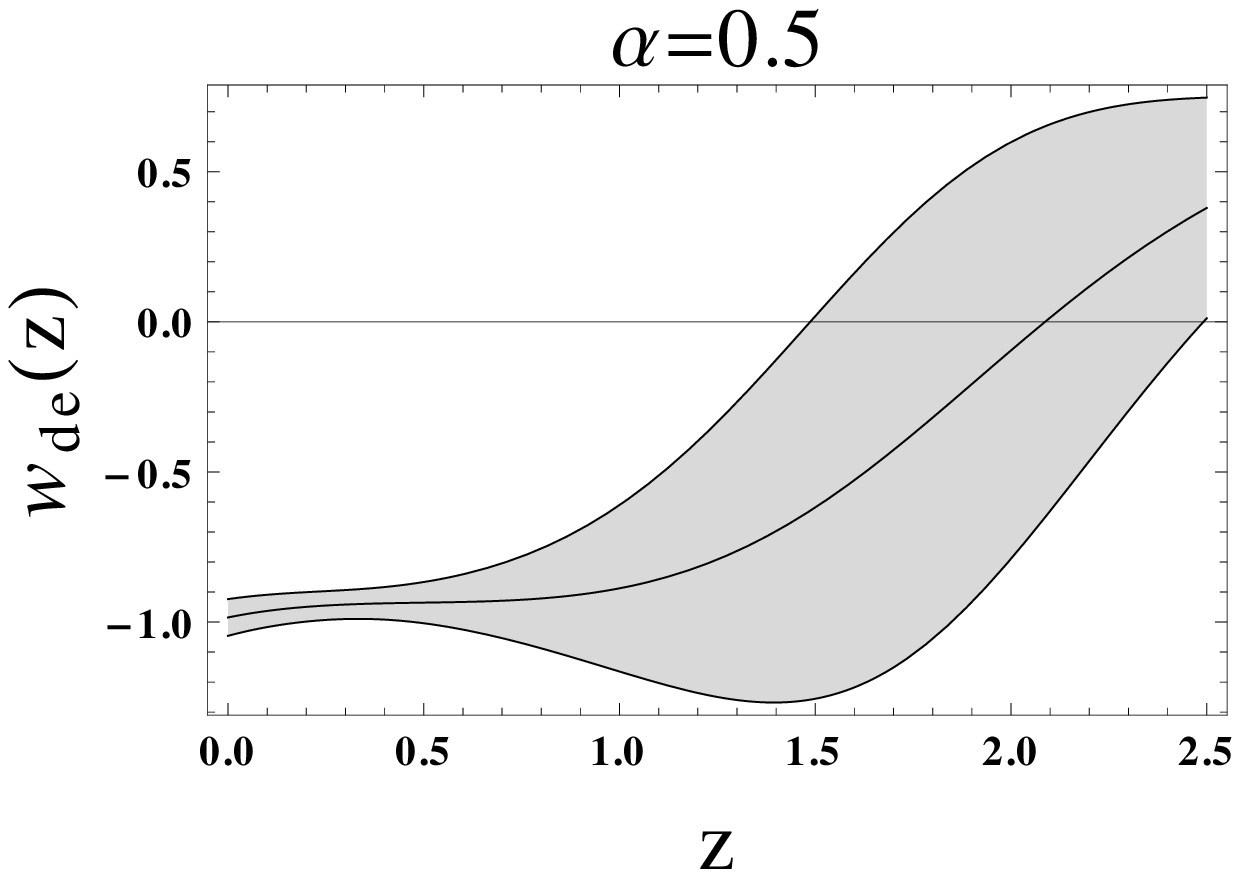}}}\hspace{3mm}
\resizebox{4.2cm}{!}{\rotatebox{0}{\includegraphics{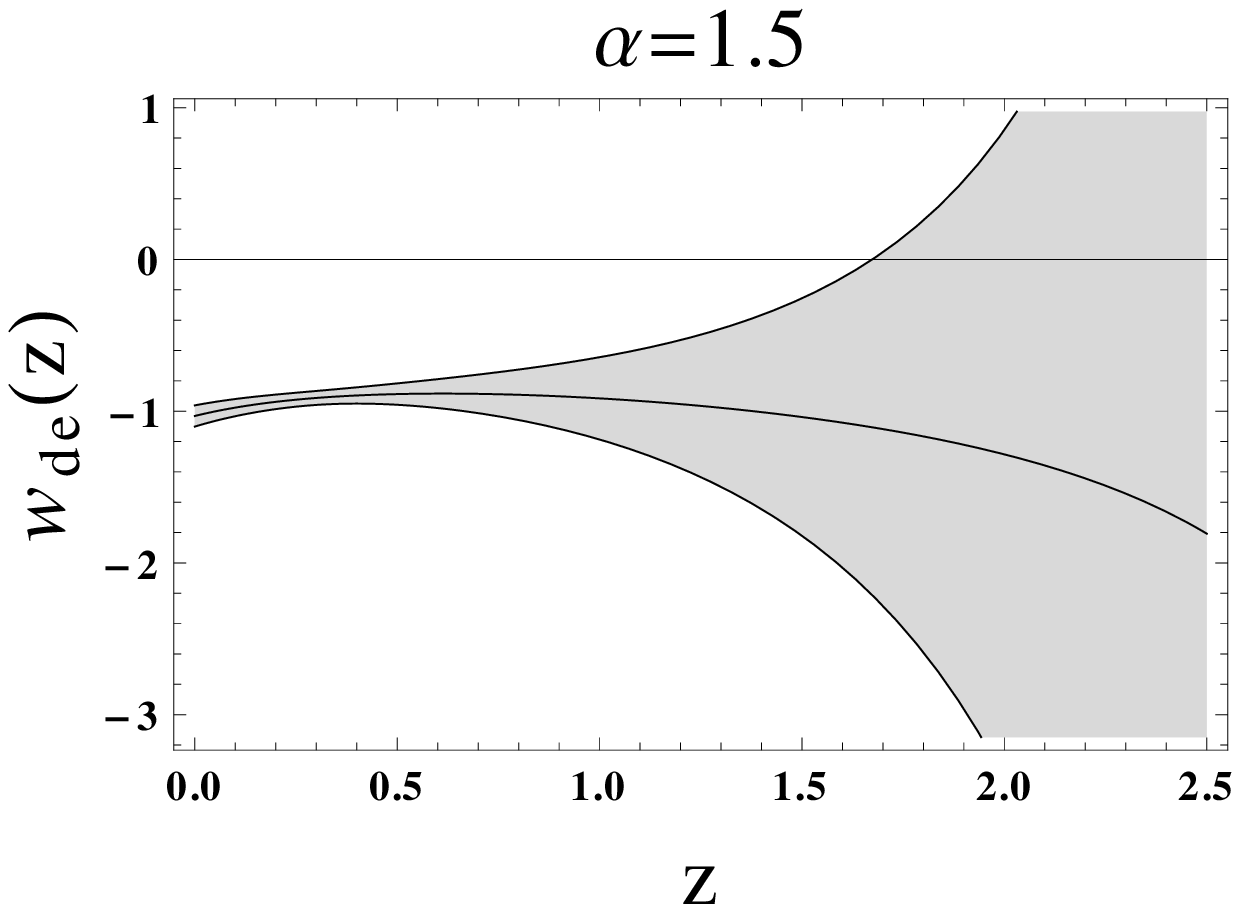}}}%
\caption{The evolution of $w_{de}(z)$ is shown for the SNIa+CC+$H_{0}$ dataset by considering different values of $\alpha$, as indicated in each panel. In each panel, the central thin line and the light gray region represent the best fit curve and the $1\sigma$ confidence level respectively. All the plots are for $\Omega_{m0}=0.3$.}
  \label{figw}
\end{figure*}

 The evolution of the deceleration parameter $q(z)$ with $1\sigma$ error is shown in figure \ref{figq} by considering different values of $\alpha$. The reconstruction of $q(z)$ is done by the combined (SNIa+CC+$H_{0}$) dataset. It has been found that for the present model, $q(z)$ shows a signature flip at the transition redshift $z_{t}=0.69^{+0.09}_{-0.06}$, $0.65^{+0.10}_{-0.07}$, and $0.61^{+0.12}_{-0.08}$ within $1\sigma$ errors for $\alpha=-0.5,0.5$ and $1.5$ respectively. This is well consistent with the previous results given in \citep{fr2013,maga2014,nair2012,mamon2016a,mamon2017}, which is also in good agreement with the $\Lambda$CDM value ($z_{t}\approx 0.7$). Similarly, the evolution of $w_{de}(z)$ with $1\sigma$ error is shown in figure \ref{figw} for the (SNIa+CC+$H_{0}$) dataset, by considering different values of $\alpha$. It is observed from figure \ref{figw} that the present value of $w_{de}(z)$ with $1\sigma$ error is very close to $-1$ in all cases (i.e., $w_{de}(z=0)=-0.94^{+0.05}_{-0.05}, -0.99^{+0.06}_{-0.06}$ and $-1.02^{+0.07}_{-0.07}$ for $\alpha=-0.5,0.5$ and $1.5$ respectively). Therefore, the results are consistent with the $\Lambda$CDM model. It is also observed from figure \ref{figw} that $w_{de}(z)$ becomes positive (within $1\sigma$ error) at high redshifts, which represents the early decelerated (matter dominated) phase of the universe. Thus $w_{de}(z)$ can easily accommodate both the phases of cosmic evolution, i.e., early decelerated phase and late-time accelerated phase.
\begin{table*}
\caption{Best fit values of the model parameters ($q_{0}$ and $q_{1}$) by considering different values of $\alpha$. For this analysis, we have considered CC+$H_{0}$ dataset.}
\begin{center}
\begin{tabular}{|c|c|c|c|c|c|c|}
\hline
$\alpha$&$q_{0}$&$q_{1}$& Constraints on $q_{0}$ and $q_{1}$&${\bar{\chi}}^{2}$\\
&&&(within $1\sigma$ C.L.)&\\
\hline
$-1$&$-0.39$&$0.54$&$-0.56\le q_{0} \le -0.20$, $0.18\le q_{1} \le 0.86$&$0.52$\\
\hline
$-0.5$&$-0.41$&$0.70$&$-0.60\le q_{0} \le -0.21$, $0.24\le q_{1} \le 1.08$&$0.512$\\
\hline
$0.5$&$-0.48$&$1.10$&$-0.70\le q_{0} \le -0.24$, $0.39\le q_{1} \le 1.74$&$0.507$\\
\hline
$1$&$-0.50$&$1.35$&$-0.75\le q_{0} \le -0.25$, $0.50\le q_{1} \le 2.19$&$0.505$\\
\hline
$1.5$&$-0.54$&$1.63$&$-0.80\le q_{0} \le -0.27$, $0.56\le q_{1} \le 2.61$&$0.504$\\
\hline
\end{tabular}
\label{table1}
\end{center}
\end{table*}
\begin{table*}
\caption{Best fit values of the model parameters for the analysis of SNIa dataset by considering different values of $\alpha$.}
\begin{center}
\begin{tabular}{|c|c|c|c|c|c|c|}
\hline
$\alpha$&$q_{0}$&$q_{1}$& Constraints on $q_{0}$ and $q_{1}$&${\bar{\chi}}^{2}$\\
&&&(within $1\sigma$ C.L.)&\\
\hline
$-1$&$-0.42$&$0.32$&$-0.59\le q_{0} \le -0.26$, $-0.35\le q_{1} \le 1$&$1.132$\\
\hline
$-0.5$&$-0.43$&$0.38$&$-0.61\le q_{0} \le -0.24$, $-0.47\le q_{1} \le 1.13$&$1.131$\\
\hline
$0.5$&$-0.45$&$0.52$&$-0.65\le q_{1} \le -0.25$, $-0.50\le q_{1} \le 1.55$&$1.129$\\
\hline
$1$&$-0.46$&$0.61$&$-0.67\le q_{1} \le -0.24$, $-0.57\le q_{1} \le 1.81$&$1.128$\\
\hline
$1.5$&$-0.47$&$0.70$&$-0.68\le q_{0} \le -0.23$, $-0.67\le q_{1} \le 2.02$&$1.127$\\
\hline
\end{tabular}
\label{table2}
\end{center}
\end{table*}
\begin{table*}
\caption{Best fit values of the model parameters by considering different values of $\alpha$. This is for the SNIa+CC+$H_{0}$ dataset.}
\begin{center}
\begin{tabular}{|c|c|c|c|c|c|c|}
\hline
$\alpha$&$q_{0}$&$q_{1}$& Constraints on $q_{0}$ and $q_{1}$&${\bar{\chi}}^{2}$\\
&&&(within $1\sigma$ C.L.)&\\
\hline
$-1$&$-0.46$&$0.64$&$-0.56\le q_{0} \le -0.35$, $0.38\le q_{1} \le 0.86$&$0.822$\\
\hline
$-0.5$&$-0.48$&$0.81$&$-0.60\le q_{0} \le -0.37$, $0.49\le q_{1} \le 1.08$&$0.821$\\
\hline
$0.5$&$-0.53$&$1.20$&$-0.66\le q_{0} \le -0.40$, $0.73\le q_{1} \le 1.64$&$0.821$\\
\hline
$1$&$-0.55$&$1.43$&$-0.69\le q_{0} \le -0.41$, $0.83\le q_{1} \le 1.95$&$0.823$\\
\hline
$1.5$&$-0.58$&$1.69$&$-0.71\le q_{0} \le -0.43$, $1.01\le q_{1} \le 2.26$&$0.825$\\
\hline
\end{tabular}
\label{table3}
\end{center}
\end{table*}
\section{Conclusions}\label{conclusion}
In this paper, a parametric reconstruction of the deceleration parameter has been presented. The functional form of $q$ is chosen in such a way that it reproduces three well-known $q$-parametrized models for $\alpha=\pm 1$ and $\alpha\rightarrow 0$. The advantage of this type of parametrization is that it incorporates a wide class of viable models of cosmic evolution based on the choice of the parameter $\alpha$. We have also constrained the model parameters by $\chi^{2}$ minimization technique using the CC+$H_{0}$, SNIa, and SNIa+CC+$H_{0}$ datasets. We have shown that the constraints obtained on the parameter values by the joint (SNIa+CC+$H_{0}$) analysis are very tight as compared to the constraints obtain from the SNIa dataset and the CC+$H_{0}$ dataset independently. The corresponding results are presented in tables \ref{table1}, \ref{table2}, and \ref{table3}, while figure \ref{fig1} shows the $1\sigma$ and $2\sigma$ confidence contours for our choices of the parameter $\alpha$. We have found that $q(z)$ undergoes a smooth transition from a decelerated to an accelerated phase of expansion in the recent past. This result is essential to explain both the observed growth of structures at the early epoch and the current cosmic acceleration measurements. Also, the signature flip in $q$ (from $q>0$ to $q<0$) occurs at the redshift $z_{t}=0.69^{+0.09}_{-0.06}$, $0.65^{+0.10}_{-0.07}$, and $0.61^{+0.12}_{-0.08}$ within $1\sigma$ errors for $\alpha=-0.5$, $0.5$, and $1.5$ respectively which are found to be well consistent with previous results \citep{fr2013,maga2014,nair2012,mamon2016a,mamon2017}, including the $\Lambda$CDM prediction ($z_{t}\approx 0.7$). We therefore conclude that the present parameterized model provides the values for $z_{t}$, $q_{0}$ and $w_{de}(z=0)$, which are in good agreement with recent observations for a wide range in the values of $\alpha$. 
\newpage
\section{Acknowledgments}
The author is thankful to Sudipta Das and Subhajit Saha for useful comments, and to the anonymous referee for valuable advices. The author acknowledges Science and Engineering Research Board (SERB), Govt. of India for the financial support through National Post-Doctoral Fellowship Scheme [File No:
PDF/2017/000308].  \\


\begin{thebibliography}{99}
\bibitem[Perlmutter et al. (1999)]{perl1999}Perlmutter S. et al., Astrophys. J., {\bf 517}, 565 (1999).
\bibitem[Riess et al. (1998)]{riess1998}Riess A. G. et al., Astron. J., {\bf 116}, 1009 (1998).
\bibitem[Ade et al. (2014a)]{pl2014a} Ade P. A. R. et al. [Planck Collaboration], Phys. Rev. Lett. {\bf 112}, 241101 (2014).
\bibitem[Ade et al. (2015)]{pl2015} Ade P. A. R. et al. [Planck Collaboration], Phys. Rev. Lett. {\bf 114}, 101301 (2015).
\bibitem[Ade et al. (2016a)]{pl2016a} Ade P. A. R. et al. [Planck Collaboration], Phys. Rev. Lett. {\bf 116}, 031302 (2016).
\bibitem[Ade et al. (2016b)]{pl2016b} Ade P. A. R. et al. [Planck Collaboration], A\&A, {\bf 594}, A13 (2016).
\bibitem[Ade et al. (2016c)]{pl2016c} Ade P. A. R. et al. [Planck Collaboration], A\&A, {\bf 594}, A20 (2016).
\bibitem[Eisenstein (2005)]{eisen2005}Eisenstein D. J., Astrophys. J., {\bf 633}, 560 (2005).
\bibitem[Hinshaw et al. (2013)]{hin2013}Hinshaw G. et al., Astrophys. J. Suppl., {\bf 208}, 19 (2013).
\bibitem[Komatsu et al. (2011)]{koma2011}Komatsu E. et al., Astrophys. J. Suppl., {\bf 192}, 18 (2011).
\bibitem[Seljak et al. (2005)]{seljak2005}Seljak U. et al., Phys. Rev. D., {\bf 71}, 103515 (2005).
\bibitem[Tegmark et al. (2004)]{teg2004}Tegmark M. et al., Phys. Rev. D., {\bf 69}, 103501 (2004).
\bibitem[Ade et al. (2014b)]{pl2014b} Ade P. A. R. et al. [Planck Collaboration], A \& A, {\bf 571}, A16 (2014).
\bibitem[Ade et al. (2014c)]{pl2014c} Ade P. A. R. et al. [Planck Collaboration],  A \& A, {\bf 571}, A24 (2014).
\bibitem[Ade et al. (2014d)]{pl2014d} Ade P. A. R. et al. [Planck Collaboration], A \& A, {\bf 571}, A22 (2014).
\bibitem[Copeland, Sami \& Tsujikawa (2006)]{cope2006}Copeland E. J., Sami M., Tsujikawa S. Int. J. Mod. Phys. D, {\bf 15}, 1753 (2006).
\bibitem[Martin (2008)]{martin2008}Martin J., Mod. Phys. Lett. A, {\bf 23}, 1252 (2008).
\bibitem[Sahni \& Starobinsky (2000)]{vs2000}Sahni V., Starobinsky A. A., Int. J. Mod. Phys. D, {\bf 9}, 373 (2000).
\bibitem[Bamba et al. (2012)]{bamba2012} Bamba K., Capozziello S., Nojiri S., Odintsov S. D.,
Astrophys. Space Sci. {\bf 342}, 155 (2012).
\bibitem[Peebles \& Ratra (2003)]{peeb2003}Peebles P. J. E., Ratra B., Rev. Mod. Phys., {\bf 75}, 559 (2003).
\bibitem[Steinhardt et al. (1999)]{Steinhardt1999}Steinhardt P. J., et al.,  Phys. Rev. Lett., {\bf 59}, 123504 (1999).
\bibitem[Weinberg (1989)]{wein1989}Weinberg S.,  Rev. Mod. Phys., {\bf 61}, 1 (1989).
\bibitem[Turner \& Riess (2002)]{turner2002}Turner M. S., Riess A. G.,  Astrophys. J., {\bf 569}, 18 (2002).
\bibitem[Aksaru et al. (2014)]{aksaru2014}Aksaru O. et al.,  EPJ Plus, {\bf 129}, 22 (2014).
\bibitem[Cunha \& Lima (2008)]{chuna2008}Cunha J. V., Lima J. A. S.,  MNRAS, {\bf 390}, 210 (2008).
\bibitem[Cunha (2009)]{chuna2009}Cunha J. V.,  Phys. Rev. D, {\bf 79}, 047301 (2009).
\bibitem[del Campo et al. (2012)]{del2012}del Campo S. et al.,  Phys. Rev. D, {\bf 86}, 083509 (2012).
\bibitem[Nair et al. (2012)]{nair2012}Nair R. et al.,  JCAP, {\bf 01}, 018 (2012).
\bibitem[Santos, Carvalho \& Alcaniz (2011)]{santos2011}Santos B., Carvalho J. C., Alcaniz J. S.,  Astropart. Phys., {\bf 35}, 17 (2011).
\bibitem[Gong \& Wang (2006)]{gong2006}Gong Y., Wang A.,  Phys. Rev. D., {\bf 73}, 083506 (2006).
\bibitem[Gong \& Wang (2007)]{gong2007}Gong Y., Wang A., Phys. Rev. D., {\bf 76}, 043520 (2007).
\bibitem[Mamon \& Das (2016a)]{mamon2016a}Mamon A. A., Das S.,  Int. J. Mod. Phys. D., {\bf 25}, 1650032 (2016).
\bibitem[Mamon \& Das (2016b)]{mamon2016b}Mamon A. A., Das S., Eur. Phys. J. C, {\bf 77}, 495 (2017).
\bibitem[Riess et al. (2004)]{riess2004}Riess A. G., et al.,  Astrophys. J., {\bf 607}, 665 (2004).
\bibitem[Xu \& Liu (2008)]{xu2008}Xu L., Liu H., Mod. Phys. Lett. A, {\bf 23}, 1939 (2008).
\bibitem[Xu \& Lu (2009)]{xu2009}Xu L., Lu J.,  Mod. Phys. Lett. A, {\bf 24}, 369 (2009).
\bibitem[Barboza et al. (2009)]{motibeosp}Barboza E. M.,  Alcaniz J. S.,  Zhu Z. -H., Silva R., Phys. Rev. D, {\bf 80}, 043521 (2009).
\bibitem[Jimenez \& Loeb (2002)]{jim2002}Jimenez R., Loeb A.,  Astrophys. J., {\bf 573}, 37 (2002).
\bibitem[Moresco et al. (2016)]{more2016}Moresco M. et al.,  JCAP, {\bf 05}, 014 (2016).
\bibitem[Moresco (2015)]{more2015}Moresco M.,  MNRAS, {\bf 450}, L16 (2015).
\bibitem[Simon, Verde \& Jimenez (2005)]{simon2005}Simon J., Verde L., Jimenez R.,  Phys. Rev. D., {\bf 71}, 123001 (2005).
\bibitem[Stern et al. (2010)]{stern2010}Stern D. et al.,  JCAP, {\bf 02}, 008 (2010).
\bibitem[Zhang et al. (2014)]{zhang2014}Zhang C. et al.,  Res. Astron. Astrophys., {\bf 14}, 1221 (2014). 
\bibitem[Betoule et al. (2014)]{Betoule2014}Betoule M. et al.,  A\& A, {\bf 568}, A22 (2014).
\bibitem[Farooq, Mania \& Ratra (2013)]{fmr2013}Farooq O., Mania D., Ratra B.,  Astrophys. J., {\bf 764}, 138 (2013).
\bibitem[Aghamohammadi et al. (2013)]{refapss1} Aghamohammadi A., Saaidi K., Mohammadi A., Sheikhahmadi H., Golanbari T., Rabiei S. W., Astrophys. Space. Sci. {\bf 345}, 17 (2013).
\bibitem[Rabiei1 et al. (2016)]{refepjc2} Rabiei1 S. W., Sheikhahmadi H., Saaidi K., Aghamohammadi A., Eur. Phys. J. C, {\bf 76}, 66 (2016).
\bibitem[Saaidi et al. (2012)]{refijmpd3} Saaidi K., Aghamohammadi A., Sabet B., Farooq O., Int. J. Mod. Phys. D, {\bf 21}, 1250057 (2012).
\bibitem[Farooq \& Ratra (2013)]{fr2013}Farooq O., Ratra B.,  Astrophys. J., {\bf 766}, L7 (2013).
\bibitem[Magana et al. (2014)]{maga2014}Magana J. et al.,  JCAP, {\bf 10}, 017 (2014).
\bibitem[Mamon, Bamba \& Das (2017)]{mamon2017}Mamon A. A., Bamba K., Das S., Eur. Phys. J. C, {\bf 77}, 29 (2017).
\end{thebibliography}
\end{document}